# Naturally Supervised Learning in Manipulable Technologies

Bradly Alicea
bradly.alicea@ieee.org
Department of Animal Science, Michigan State University

Keywords: Biomechanics, Brain-Computer Interfaces, Virtual Reality

## ABSTRACT

The relationship between physiological systems and modern electromechanical technologies is fast becoming intimate with high degrees of complex interaction. It can be argued that muscular function, limb movements, and touch perception serve supervisory functions for movement control in motion and touch-based (e.g. manipulable) devices/interfaces and human-machine interfaces in general. To get at this hypothesis requires the use of novel techniques and analyses which demonstrate the multifaceted and regulatory role of adaptive physiological processes in these interactions. Neuromechanics is an approach that unifies the role of physiological function, motor performance, and environmental effects in determining human performance. A neuromechanical perspective will be used to explain the effect of environmental fluctuations on supervisory mechanisms, which leads to adaptive physiological responses. Three experiments are presented using two different types of virtual environment that allowed for selective switching between two sets of environmental forces. This switching was done in various ways to maximize the variety of results. Electromyography (EMG) and kinematic information contributed to the development of human performance-related measures. Both descriptive and specialized analyses were conducted: peak amplitude analysis, loop trace analysis, and the analysis of unmatched muscle power. Results presented here provide a window into performance under a range of conditions. These analyses also demonstrated myriad consequences for force-related fluctuations on dynamic physiological regulation. The findings presented here could be applied to the dynamic control of touch-based and movement-sensitive human-machine systems. In particular, the design of systems such as human-robotic systems, touch screen devices, and rehabilitative technologies could benefit from this research.

## INTRODUCTION

In the last few years, there have been major advances in the commercial availability of touch-based and motion-driven devices. Devices such as the iPad, Nintendo Wii, Microsoft Kinect, and other applications have required a new way of thinking about usability and ergonomic design. Parallel developments in the area of brain-computer interfaces [1, 2], neurorehabilitation [3, 4], and myoelectric control [5] may provide clues to this issue, but have not yet become a key component of human factors research. In this paper, I will demonstrate how an approach called neuromechanics [6, 7] can be brought to bear on assessing the usability of such technologies. Neuromechanics is an approach that unifies the complexity of behavior with neurobiological outputs by looking at interactions between muscle activity and movement behavior. Examples include dynamical control strategies due to limb geometry [8] and neural mechanisms [9], which can be characterized as morphological and neural control strategies, respectively.

**Neuromechanical control as supervised learning**
The human neuromechanical system can be understood as a complex control system which can adapt to environmental stimuli. Experiments in motor learning [10] and computational neurobiology [11] have demonstrated the role of switching between forces in the learning of complex movements. In particular, the ability to recover from experimentally-induced distortions serves as a buffering mechanism for dealing with changes in velocity and acceleration that occur during the execution of complex movements. For example, the healthy movement system is generally robust to both incremental and nonlinear fluctuations in inertial and rotational forces when performing real-world tasks [12, 13]. For example, muscles can act as both energy stores and shock absorbers [14]. By introducing such force field distortions during a training period, the neuromechanical system adapts to fluctuating conditions by a mechanism I am calling biological supervised learning.

The concept of supervised learning [15] is widely used in artificial intelligence [16], particularly in the design of movement controllers for robots. While the biological analogues of this control strategy are not completely understood, we may utilize this conceptual model as a means to achieve improved performance in human-machine interaction. By using explicitly physical distortions of force in conjunction with the movement of simulated objects, an experimental session can be tightly controlled without losing the naturalistic aspects of this supervisory mechanism.

Unlike the top-down exemplars of categories typical in artificial supervised learning [17], the proposed natural supervisory mechanism involves the interactions of environmental structure and proprioception. This is mediated by morphological control [18], which involves coordination of the limbs and muscle activity during movement. While this does not provide a dominant bottom-up mechanism observed in self-organized behaviors [19], the proposed supervised learning mechanism does allow for an overall more adaptable response that maximizes functionality of human-machine (hybrid physiological) systems.

**Research Questions and Assumptions**
This rationale for this work is related to previous work on human performance augmentation [20], Brain-computer Interfaces [21], and Augmented Cognition [22]. For the latter two applications in particular, tightly-integrated closed-loop control is essential for minimizing system error. While this work does not utilize a closed-loop system, the findings are relevant to maintaining closed-loop control. Therefore, this work fills a gap in the literature by way of addressing two questions. One question involves what physiological/cognitive responses and how much variability characterizes internal mechanisms that serve as responses to environmental stimuli. The other question involves the supervisory nature of environmental stimuli themselves, particularly when they are variable or switch-like. A related rationale involves working towards design principles for health-care and rehabilitative applications. In these cases, this work should provide two useful outcomes: understanding the ability of clinical populations to recover from environmental distortions, and how training mechanisms can be implemented in virtual environments.

To answer both questions and place it in the context of technological applications, the experimental investigations presented here featured the following: sets of environmental forces imposed on the upper limbs during reaching and active touch exploration, and presenting alternating sets of forces in sequences of variable length. This is similar to the learn-unlearn-relearn experimental approach derived from the motor learning [23] and aerospace medicine [24] communities. While it may appear that introducing alternate conditions at different points in the learning sequence is a confounding variable, it is also worth noting that from a naturalistic standpoint, the phenomena of stimulus presentation order and timing of switching are not independent. Because we do not always associate a switch in environmental conditions with either learning or unlearning provides us with a series of internal controls through which we can investigate contextual effects. In fact, sequence of presentation is perhaps the most important factor for uncovering the effects of switching and function of the hypothesized internal mechanisms. The driving force behind the expected changes involve switching between different force fields, which is approximated in these experiments by physically loading the hand and/or arm with different types of forces.

**Manipulations and Measures**

In these experiments, environmental structure is introduced using a virtual simulation with limited force feedback. In two of the three experiments, switching is introduced using a tool with a variable weight on its distal end. This provides an unstable radius of gyration for an otherwise deterministic task. In the third experiment, switching is introduced by the virtual environment itself though variable resistance during the free exploration of different surface types. This provides an unstable elasticity for an otherwise deterministic task.

In this paper, inertial and/or rotational forces will be used to perturb haptic (e.g. the sense of touch) or truly proprioceptive (e.g. the sensation of joint and limb movement) sensory inputs in semi-natural contexts. The contributions of haptic/proprioceptive sensation and learning to natural supervised learning can be measured indirectly using the mapped physiological output (MPO) and unmatched muscle power (UMP) variables. These are dependent on two interrelated environmental parameters: inertial feedback and the strength of force distortions, which contribute to dynamic responses exhibited by the hybrid physiological system.

There is also a relationship between environmental forces that serve as sensory input and muscle power that serves as physiological and behavioral output. The concept of muscle power [6, 25] can be used to understand the role of unmatched muscle power. Traditionally, a measure of muscle power involves assessing the role of muscle length, particularly the shortening velocity of a particular muscle during movement, to produce a force output. Muscle power can be thought of as energy per unit of object movement. When a discrepancy exists between electrophysiological amplitude and the movements of objects in a virtual environment, then unmatched muscle power will result. Observing changes in unmatched muscle power, particularly across conditions, is particularly informative for understanding the effects of switching between forces.

## METHODS

**Participants**

Thirty-seven participants were used in the swinging device switching experiment (Experiment #1). Thirty-two subjects were used in the extended swinging device switching experiment (Experiment #2). Fifteen subjects were used in the tactile surface switching experiment (Experiment #3). Participants gave their informed consent before participating in the study. The study was approved by the local ethics committee and performed in accordance with local ethics committee standards. Participants gave their informed consent before participating in the study. The study was approved by the local ethics committee and performed in accordance with local ethics committee standards.

**Design for Experiment #1: Swinging Device Switching Experiment**

An experiment called the swinging device switching experiment was conducted to better understand the immediate effects of haptic/proprioceptive switching on a short-term training regimen. A 2 (swinging device) x 3 (learning blocks) x 4 (switching type) x 16 (trials) mixed experimental design was used. The between subjects factors are swinging device and switching, while the within subjects factors are learning blocks and trials. Swinging device setting has two levels: unloaded controller and loaded controller. Switching type has two levels: switching and no switching. Learning type has three levels, learning, unlearning, and relearning. Table 1 shows the specifics of the 2x3x4x16 design.

**Table 1. Experiment #1: 2x3x4x16 factorial design for the prosthetic device switching experiment. L = loaded, U = unloaded.**

| Sequence Presentation | L, U, L | U, L, U | L, U, U | U, L, L |
|---|---|---|---|---|
| **Learning Block 1** | 16 trials | 16 trials | 16 trials | 16 trials |
| **Learning Block 2** | 16 trials | 16 trials | 16 trials | 16 trials |
| **Learning Block 3** | 16 trials | 16 trials | 16 trials | 16 trials |

**Design for Experiment #2: Extended Swinging Device Switching Experiment**

The extended swinging device switching experiment was conducted to better understand the extended effects of haptic/proprioceptive switching on a short-term training regimen. A 2 (swinging device) x 4 (switching type) x 5 (learning blocks) x 16 (trials) mixed experimental design was used. The between subjects factors are swinging device and switching, while the within subjects factors are learning blocks and trials. Swinging device setting has two levels: unloaded controller and loaded controller. Switching type has four levels: interleaved, early, late, and none (control). Learning type has five levels, learning, unlearning, secondary learning, secondary unlearning, and tertiary learning. Table 2 shows the specifics of the 2x4x5x16 design.

**Table 2. Experiment #2: 2x4x5x16 factorial design for the extended prosthetic device switching experiment. All switching is achieved using the prosthetic device (loaded condition).**

| Sequence Presentation | Alternate* condition | Early condition | Late condition | None (control) condition |
|---|---|---|---|---|
| **Learning** | 16 trials | 16 trials | 16 trials | 16 trials |
| **Unlearning** | 16 trials | 16 trials | 16 trials | 16 trials |
| **Secondary Learning** | 16 trials | 16 trials | 16 trials | 16 trials |
| **Secondary Unlearning** | 16 trials | 16 trials | 16 trials | 16 trials |
| **Tertiary Learning** | 16 trials | 16 trials | 16 trials | 16 trials |

* alternate = switching during learning, secondary learning, and tertiary learning (blocks 1, 3, and 5).

**Design for Experiment #3: Tactile Surface Switching Experiment**
The tactile surface switching experiment was conducted to assess whether or not effects similar to those observed in Experiments #1 and #2 were also observed specifically in the context of touch. Experiments #1 and #2 involved introducing and removing surface reaction forces typical of a swinging device with fixed physical parameters. An alteration between loading and unloading the arm with rotational forces resulted in an effect on task performance. It was therefore suspected that variable inertial forces from different surfaces would demonstrate various types of effects demonstrated by the swinging device switching experiment. As a result, a 3(switching type) x 3(block) mixed experimental design was conducted to investigate the effects of free exploration of different surface types on muscle activity. To more directly address the effect of differential loading, switching is introduced at a uniform point in the sequence. The between-subject factor was switching type. Switching has three levels: hard switching, weak switching, and reverse switching. Block had three levels: surface learning block 1, surface learning block 2, and surface learning block 3. Table 3 shows the specifics of this 3x3 design.

**Apparatus Protocol**
This section describes various features of the apparatus used in these experiments. This includes the instrumentation used to produce measures and manipulate performance.

*Apparatus for Experiments #1 and #2.* The simulation portion of both swinging device switching experiments will be conducted using the Nintendo® Wii gaming platform (Nintendo Corporation, Kyoto, Japan) using the simulation *Wii Sports Golf* (Figure 1). The Nintendo® Wii uses several simulation simulation-specific parameters and a motion controller to produce computer-generated simulation action. All simulation-specific parameters, such as wind speed and club surface type, will be held constant.

**Table 3. Experiment #2: 3x3 factorial design for the tactile surface switching experiment. All surface names are defined as they are in the Novint tutorial program.**

| Switching Type | **Surface Exploration Block 1** | **Surface Exploration Block 2*** | **Surface Exploration Block 3** |
|---|---|---|---|
| **Hard** | Magnetic (3 minutes) | Honey (3 minutes) | Ice (3 minutes) |
| **Weak** | Bumpy (3 minutes) | Rubber (3 minutes) | Sandpaper (3 minutes) |
| **Reverse** | Honey (3 minutes) | Bumpy (3 minutes) | Sand (3 minutes) |

* switching introduced.

*Apparatus for Experiment #3.* The simulation portion of the tactile surface switching experiment will be conducted using the Novint Falcon (Novint Corporation, Albuquerque, NM) 6 degree-of-freedom force-feedback device (Figure 2). This input/output controller was used to manipulate a virtual sphere tiled with various surfaces (Figure 2). During the experiment, the subject manipulated a knob which moved the arms of the controller. These movements were mapped to a cursor in the virtual environment, with which each simulated surface was freely explored.

Switching will be achieved using two mechanisms: a distortion of forces during a commonly encountered repetitive task that subjects were allowed to informally practice on beforehand, and a quick, drastic changes in the forces being explored for 3 minute intervals.

*Switching for Experiments #1 and #2.* For both swinging device switching experiments, the purpose of the swinging device is to simulate a set of environmental conditions that perturb upper-arm morphology and the current physiological state. For both loaded and unloaded conditions, the Wiimote motion controller uses a gyroscope to translate actions produced by the human into a virtual analogue of physical movement. For conceptual comaprison, the Segway uses a similar mechanism to map user input to continuous motion. While this produces gyroscopic forces of limited scope which act as a source of inertial feedback, the gravity and velocity effects of the simulation physics should be consistent within each condition. In the loaded conditions, the addition of the swinging device forces participants to perform the putting task in the context of a distorted pendulum.

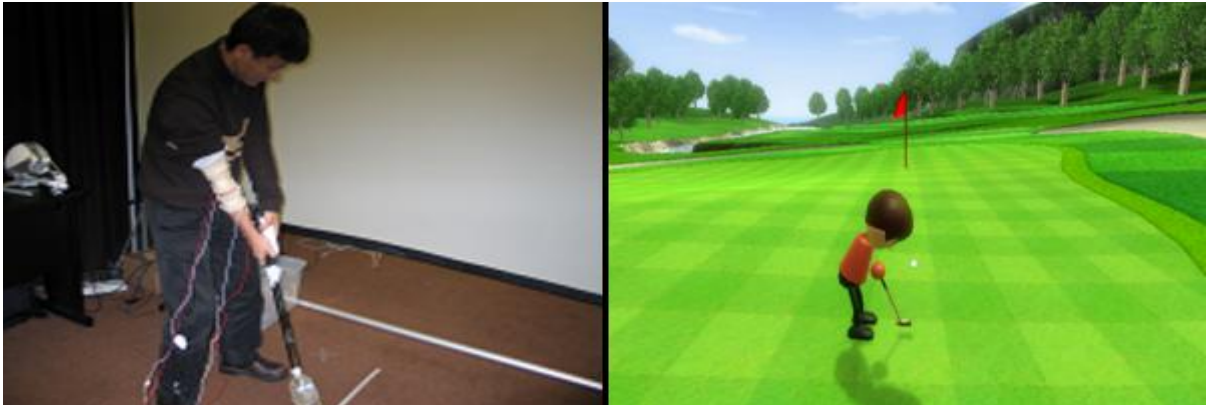

**Figure 1. Images of the experimental setup and apparatus for both swinging device switching experiments. Left: image of an individual standing in front of the CAVE-like virtual environment wearing electrodes during an example of the loaded experimental condition. Right: a screenshot of the experimental task, a Wii Sports putting simulation.**

**Experimental Switching**
*Unloaded vs. loaded conditions.* The unloaded controller conditions will involve using the controller that is standard with the Wii system. In unloaded conditions, participants manipulate action in the virtual environment with a motion controller alone, mimicking the motions of the reaching activity without any of the feedback from the swinging device (i.e. motion coupling between the swinging device and counterweight).

The loaded controlled consists of a Wiimote motion controller strapped to a customized swinging device. In this experiment, the swinging device consists of two golf clubs bound together at the shaft with a dynamic counterweight (a bottle filled with a liquid of specific density) attached at the base of the shafts. The swinging device was designed this way to ensure that previous experience with golf clubs or other swinging tools was minimized and that a switching due to loading could be systematically introduced.

The loaded condition provides switching along two axes of movement input relative to the virtual environment [26] and provides a distortion of haptic/proprioceptive information that complements the absence of such information as presented by the unloaded controller. During both the unloaded and loaded blocks, the operator will make sixteen reaches with respect to the Wii sports simulation. As the task takes place on a virtual putting green, the goal of each trial will be to get the ball in the hole as often as possible. When successful, this will minimize the variance of the mapped physiological output measurement for a single trial.

*Switching for Experiment #3.* For the tactile surface switching experiments, switching was embedded in three types of sequences: hard, weak, and reverse. In all cases, the switch to the surface coincided with the second (unlearning) block. All simulated surfaces were presented as the edges of a sphere. A hard switch was defined by a honey surface in between surfaces simulating a magnet and ice. A weak switch was defined by a rubber surface in between surfaces simulating a series of bumps and sandpaper. The honey surface was referred to as a "hard" switch because honey has a higher surface resistance than rubber (which was defined as the soft switch). In the reverse switch, a really hard surface (the series of bumps) was presented in between honey and sand, which already have a relatively high surface resistance.

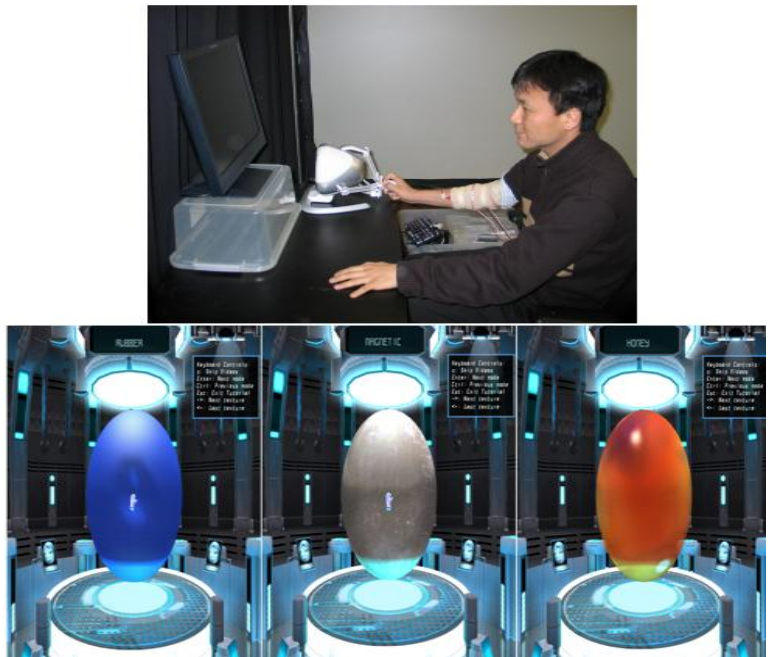

**Figure 2. Images of experimental setup and apparatus for the tactile surface switching experiment. Top: image of an individual seated in front of the Novint simulation environment wearing electrodes and engaging the Novint force-feedback device with their right hand. Bottom: three examples of surfaces presented in the Novint environment (from left: rubber, magnetic, and honey). Participants touch a sphere with these properties using a virtual hand (cursor-like object in front of the rubber and magnetic spheres).**

**Electromyography**

Surface electromyography (EMG) was collected using the Biopac MP150 amplifier. Hardware filtering was done to remove ambient noise, while an infinite-impulse response (IIR) filter was used to remove potential ECG artifacts. EMG-related activity was collected from two points on the dominant arm corresponding with two muscles: the traceps brachii (TB) and flexor carpi radialis (FCR). The triceps brachii corresponds to the humerus, while the flexor carpi radialis corresponds to the forearm.

The loop trace graphs for all experiments were produced using a filtered raw signal. Filtering was done in two steps. First, a low-pass infinite-impulse filter (IIR) of 40Hz has used to clean the raw signal. A 50-element long running average algorithm was then applied recursively across each time series. This down-sampled the raw signal to 30Hz. To apply EMG data from the tactile surface switching experiment to spikiness measure, the rectified signal for each 3 minute long experimental block was recursively partitioned into non-overlapping windows 10-seconds in duration using the peak detector function in AcqKnowledge 3.8.1 (Biopac Corporation, Goleta, CA). This algorithm produced a signal that represented all of the peaks for each window in the time-series.

*Preparations for EMG surface electrodes.* EMG measurements were collected using skin surface electrodes using the following protocol. An impedance check was run using a Checktrode MK-III (UFI, Morro Bay, CA) unit on the preparation before each set of trials to ensure calibration of the instrument. The skin was cleaned and abraided using a mixture of 70% isopropyl alcohol ($C_3H_8O$), 30% water ($H_30$), and electrode gel (Biopac Model GEL-1). Adhesive surface Ag-Ag-Cl electrodes (Biopac Model EL-503) were attached to appropriate places on the skin. Surface recording sites for each individual were determined using a standard skeletal muscle atlas, palpation, and measurement. The surface sites and portions of the electrode lead were then secured in place with athletic tape; athletic tape was wrapped several times around the body segment in question. Both of these procedures were done minimize shifting of the electrode and lead wires and to maintain impedance between the skin surface and the electrode.

## MEASURES AND EQUATIONS

**Switching definition**

Switching can be defined mathematically as

$$S = \frac{D_{end} - D_{begin}}{t} \qquad [1]$$

where $S$ is the degree of switching, $D_{end}$ is the magnitude of perturbation at the end of perturbation, $D_{begin}$ is the magnitude of perturbation at the end of perturbation, and $t$ is the duration of switching.

**Inertial Feedback**

The effects of the forcing chamber on the operator were kept constant by filling it with water at room temperature. The relationship between switching, specific density, weight, and volume is

$$S \propto d_S, \; d_S = \frac{W}{vol} \qquad [2]$$

where $S$ is the degree of switching, where $d_S$ is specific density ($d_{S\,=\,1}$), $W$ is weight, and *vol* is volume.

**Mapped Physiological Output Measure**

Mapped physiological output (MPO) is measured using the following equation:

$$\mathbf{MPO_t} = \frac{|D_{req} - D_{moved}|}{D_{req}} \quad [3]$$

where $D_{req}$ is the fixed distance a virtual object needs to be moved over a given trial, $D_{moved}$ is distance the virtual object is actually moved resulting from muscle force production captured by the input device, and $MPO_t$ is mapped physiological output for a single reach related to the presented task.

**Unmatched Muscle Power Measure**

Unmatched muscle power (UMP) is defined by the following equation:

$$\mathbf{UMP} = \frac{RP_i}{MPO_i} \quad [4]$$

where *UMP* is unmatched muscle power, *RP* is raw signal peak over a finite time interval corresponding with the duration of an experimental trial, and *MPO* is mapped physiological output for the trial corresponding to the *RP* window[1].

**Muscle Peak Amplitude Measure**

The peak amplitude is calculated using the following equation:

$$\mathbf{RP_i} = \frac{SG_{max}}{TR_t} \quad [5]$$

where *RP* is the raw signal peak over a finite time interval, *SG* is the EMG signal across the duration of that time interval, and *TR* is the duration of a single trial. For each of these windows, the signal was rectified and peak signal amplitude was calculated.

**Spikiness**

Spikiness is defined by the parameter *z* [27]:

$$\mathbf{z} = \frac{MAX_i - MIN_i}{\bar{x}_i} \quad [6]$$

where $MAX_i$ is the maximum value over interval *i*, $MIN_i$ is the minimum value over interval *i*, and $\bar{x}_i$ is the mean value over interval *i*.

---

[1] When the UMP measurement equals 0, there is a theoretical match between how much a given muscle is stretched during movement and the amount of resulting force mapped into the virtual environment by mechanical motion of the arm about the rotational axis of the controller. UMP measurement exceeding 0 but is less than 1 result in underpowered movement. UMP measurements exceeding 1 represents overpowered movement. A unit was defined as energy expended per meter of mapped physiological output, while energy was defined as the maximal amplitude of the EMG signal within a variable but finite time window.

## RESULTS

Basic descriptive evaluations and parametric statistical tests were conducted on the tactile surface switching experiment and extended swinging device switching experiment results. Figure 3 is a graph labeled with significant results from a Bonferroni corrected paired t-test (see Table 4 for result of all tests) for both muscles over all conditions of the tactile surface switching experiment. Significant and near-significant results between block pairs are shown using red and black brackets, respectively. Of note is the relatively greater number of significant results for a muscle representing involvement of the forearm (FCR). This suggests that active touch exploration shows direct effects in the forearm, which is evidence that activity type can influence the type of mechanism used to deal with the switching (e.g. morphological vs. physiological control).

In Figure 4, the p-values for the UMP measurement were also calculated using a paired t-test for both the triceps brachii and flexor carpi radialis muscles using data from Experiment #3. For the experimental control, or presentation of the unloaded condition during all blocks, UMP (TB) exhibits statistical significance at the .05 level when comparing blocks 2, 3, 4, and 5. The UMP measurement for flexor carpi radialis (UMP-FCR) when compared between blocks 2 and 3 are significant $p < .04$.

For the UMP (TB) measurement in the late condition, comparisons flanking the transition event between no switching and switching are significant, $p < .03$ and $p < .03$, while a comparison of blocks immediately before and after a switching point in the condition is not significant, $p > .40$. For the interleaved condition, the third switching event (from unloaded to loaded) is significant for UMP (TB), $p < .04$, and nearly significant, $p < .09$ for UMP (FCR).

**Table 4. Selected paired t-test results on TB and FCR muscles for tactile resistance switching experiment. All p-values < .05 level (Bonferroni corrected).**

| TB | Weak | Hard | Reverse |
|---|---|---|---|
| **Learning vs. Unlearning** | p > .07 * | p > .38 | p > .42 |
| **Unlearning vs. Relearning** | p > .27 | p > .18 | p > .15 |
| **Learning vs. Relearning** | p > .33 | p > .19 | p > .14 |
| FCR | Weak | Hard | Reverse |
| **Learning vs. Unlearning** | p < .01 | p > .07* | p > .10 |
| **Unlearning vs. Relearning** | p > .11 | p < .03 | p < .04 |
| **Learning vs. Relearning** | p < .02 | p > .14 | p > .47 |

* near-significance.

The proportional UMP measurements for TB and FCR (Figure 5) reveal finer detail regarding the effects of switching on the neuromechanical system. For UMP (TB), there were three distinct effects. For the interleaved condition, switching results in an immediate increase in UMP, and appears to be sustained in the tertiary learning block. The early condition shows that switching to a series of unloaded block suppresses UMP in a transient fashion, as the effect wears off in the next block. The late condition shows that UMP is also suppressed after a double dose of loaded blocks is presented.

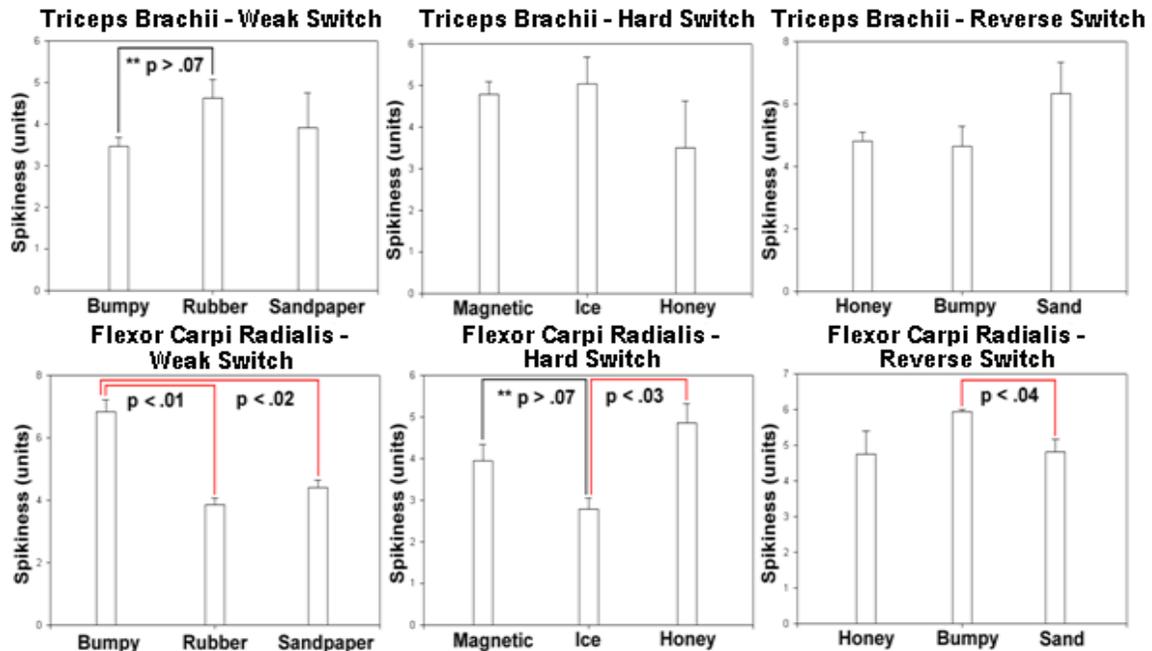

**Figure 3. Paired t-tests and significant results for the tactile resistance switching experiment (see also Table 4). Red brackets = significant results (shown). Black brackets = ** near significance**

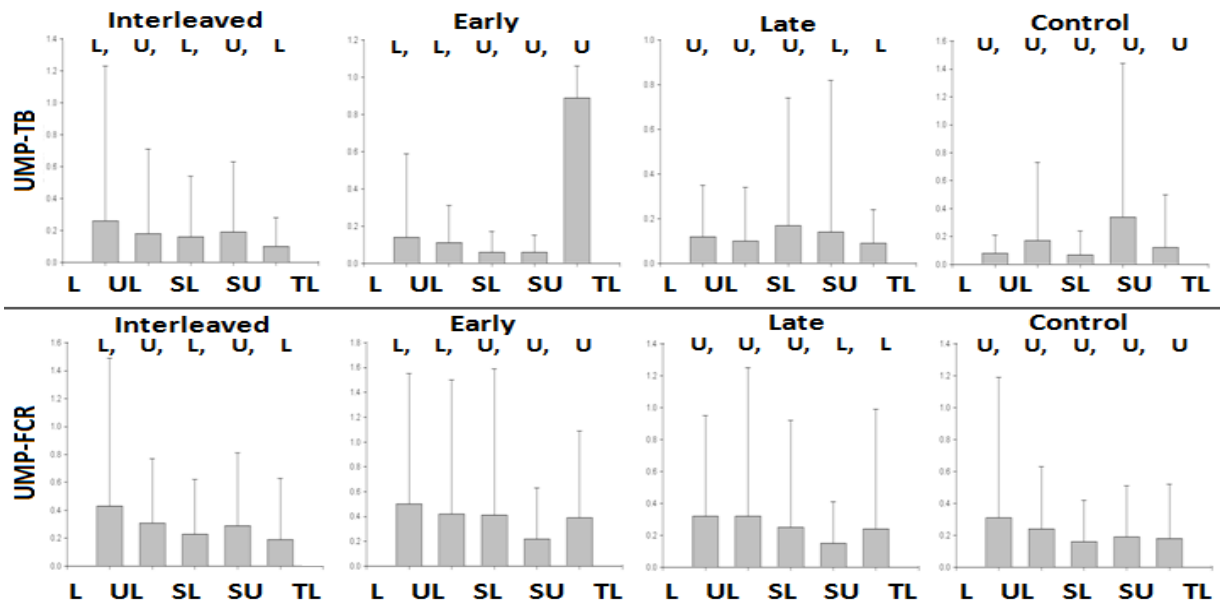

**Figure 4. Bar Charts for Unmatched Muscle Power (UMP) for Triceps Brachii and Flexor Carpi Radialis). UMP is measured in volts per meter (V · m*(0.3048)).**

The UMP (FCR) measurement shows different effects for the same condition. In the interleaved condition, UMP is stable across all blocks but is consistently lower than the control condition in all cases. For the early condition, there is a slight decrease in the response to a double dose of presenting the loaded block. In addition, there may also be a delayed response to the switching of forces. Finally, the late condition demonstrates that there is a slight decrease in UMP until a loaded block is introduced. Interestingly, secondary learning and tertiary learning have similar UMP values, which may demonstrate a transitory effect of introducing a loaded block late in the sequence.

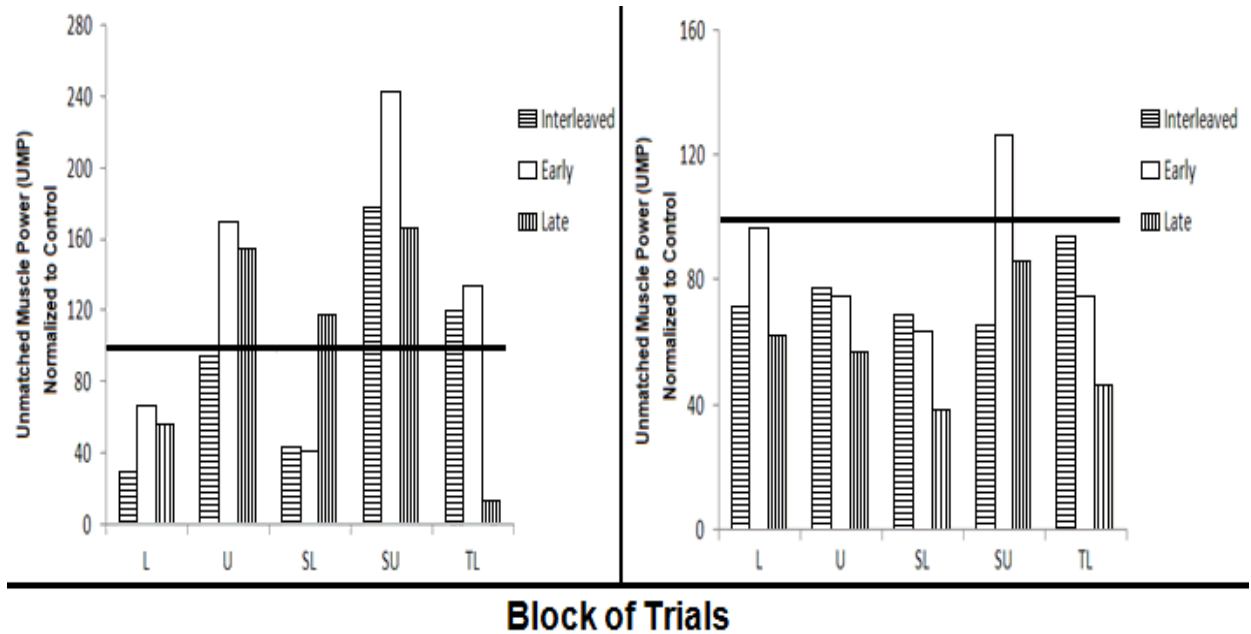

Figure 5. Unmatched muscle power (Left: Triceps Brachii (TB). Right: Flexor Carpi Radialis (FCR)) by mean across blocks of trials and as a percentage of baseline (in every instance, baseline for block is 100 and represented by black horizontal lines). UMP is measured in volts per meter (V · m*(0.3048)).

**Peak Amplitude Analysis**

To better understand how the muscle activity showed variability with respect to the different experimental conditions, I conducted a peak amplitude analysis. This analysis was conducted to answer the question of how much muscles are being activated relative to the amount of activation required. The first step was to extract the raw EMG signal recorded for each muscle. For each block, the raw signal was partitioned into 16 windows, each corresponding to a particular experimental trial. These data were averaged across all individuals who were administered a particular condition.

*Relating peak amplitude to mapped physiological output.* The next step involved comparing these peak data with the MPO measurement for each trial. This was done using the UMP measurement. Figures 1 and 3 demonstrate the relationship between these two variables for each muscle recorded. As demonstrated in Figures 3 and 4, the activity of TB and FCR is affected by both the experimental conditions of a particular block and the conditions of previous blocks. For example, the before and after switching blocks for TB (Figure 6), activity is within the same range for both unloaded and loaded blocks. For this muscle, activity increases for larger values of MPO and decreases for smaller values of MPO. When comparing the before and after switching blocks for FCR (Figure 7), there appears to be a reduction in unmatched muscle power for the unloaded blocks but an overall increase in the loaded blocks. Furthermore, there appears to be greater activity for smaller MPO values.

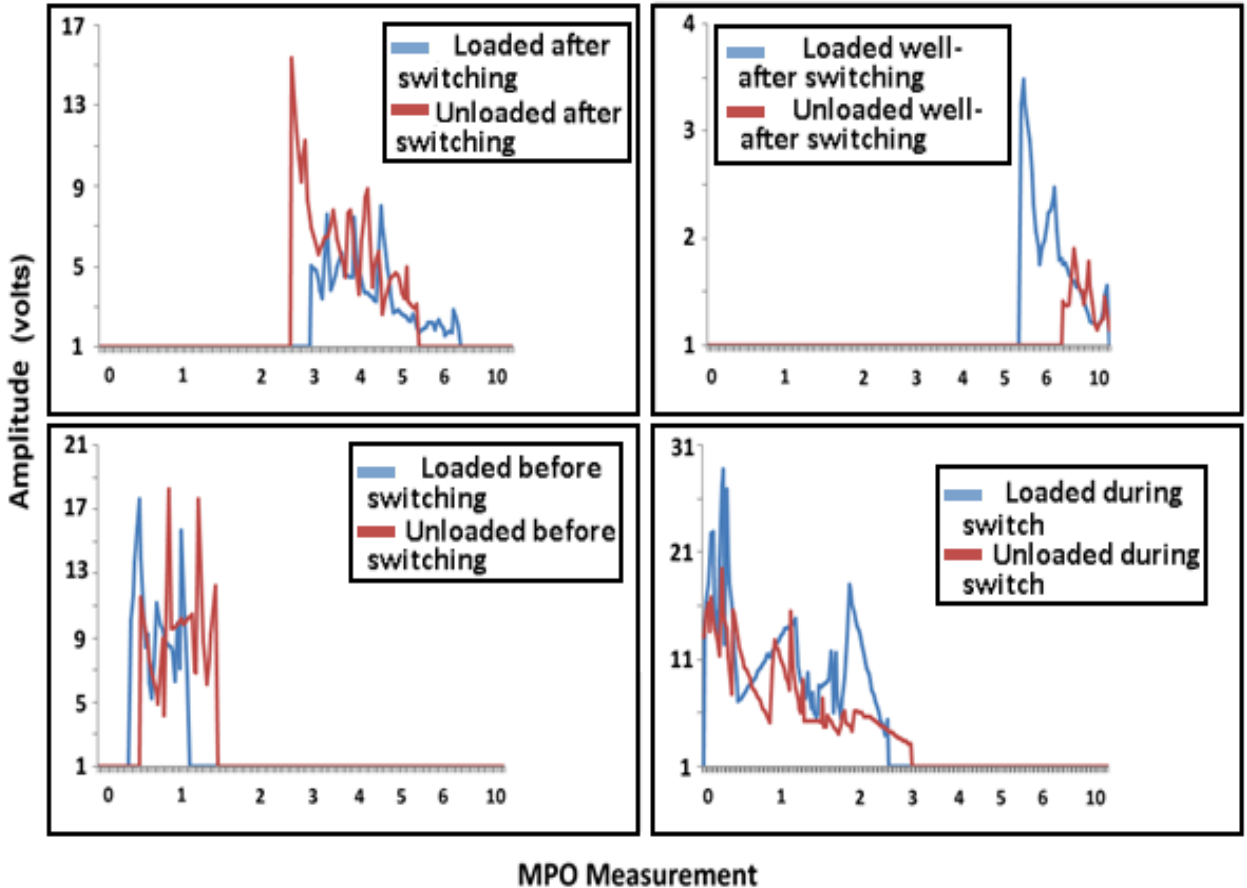

**Figure 6. Components of unmatched muscle power for Triceps Brachii (TB). Counterclockwise: Upper left, comparison of loaded after switching and unloaded after switching. Lower left, comparison of loaded before switching and unloaded before switching. Lower right, comparison of the loaded after switching and unloaded after switching. Upper right, comparison of unloaded well-after switching and loaded well-after switching.**

**Analysis of unmatched muscle power (UMP)**

The next step is to discover the relative proportion of unmatched muscle power for several classes of UMP values characterized by each experimental block. This analysis was conducted to demonstrate the mismatch of forces before and after switching and typical responses to encountered forces. Figures 8 and 9 show the results of Figures 6 and 7 in histogram form. Peaks in the histogram show the relative abundance of UMP measurements for a certain value. It was found that across both muscles, the loaded technological device resulted in a raw signal with greater amplitude before and during a loaded block, and a greater degree of unmatched muscle power produced after and well-after switching. This result was reversed somewhat for the FCR muscle during a loaded block, which may involve the predominance of muscular control in this context. This may suggest that a combination of switching and previous experience will result in a particular neuromuscular response.

*Differential effect of switching type.* For the flexor carpi radialis (FCR, Figure 8), the results can be divided into the effects of switching to loaded and unloaded blocks. Before the loaded block was introduced, 90% of the UMP values were between 3 and 4, while a very small proportion of values up to 30 were represented. After the loaded block was introduced, the majority of UMP values were under 1, while a small proportion of values were either 0 or between 1 and 1.5. This means that switching to the loaded block created a situation where muscle power was being produced in such a way that reduced the amount of force translated into the virtual environment by the controller.

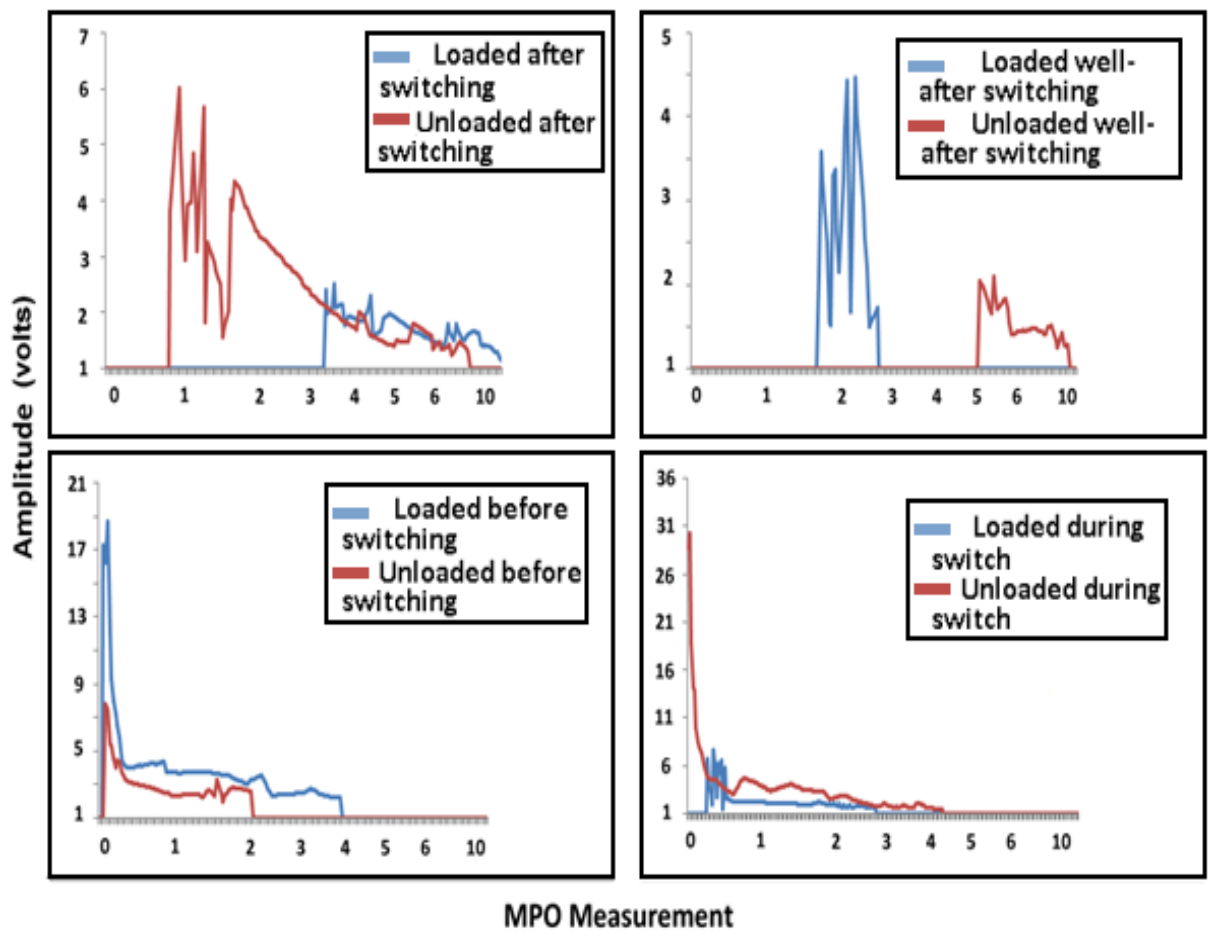

**Figure 7. Components of unmatched muscle power for Flexor Carpi Radalis (FCR). Counterclockwise: Upper left, comparison of loaded after switching and unloaded after switching. Lower left, comparison of loaded before switching and unloaded before switching. Lower right, comparison of switching to loaded block and switching to unloaded block. Upper right, comparison of unloaded well-after switching and loaded well-after switching.**

In the case of switching to the unloaded block, 75% the UMP values were distributed mainly between 1 and 3.5 before switching, while after switching a little more than half of the values were distributed between 0 and 1, while there was a broader distribution of values between 1 and 6. Again, the production of muscle power was being shifted after the unusual blocks in a sequence in a way that favored the underproduction of force.

Overall, the triceps brachii (TB - Figure 9) demonstrated similar effects for both the loaded and unloaded switches. While the TB and FCR are part of the same mechanical system that results in a physiological output, they are on different segments of the dominant arm, which makes comparisons of their contribution to force output important. After each switching type, there was a downward shift in the values of the UMP value exhibited in this population. Both before and after perturbation, the majority of UMP values were above 1. This means force production was overpowered both before and after switching, but less so after the switching was introduced.

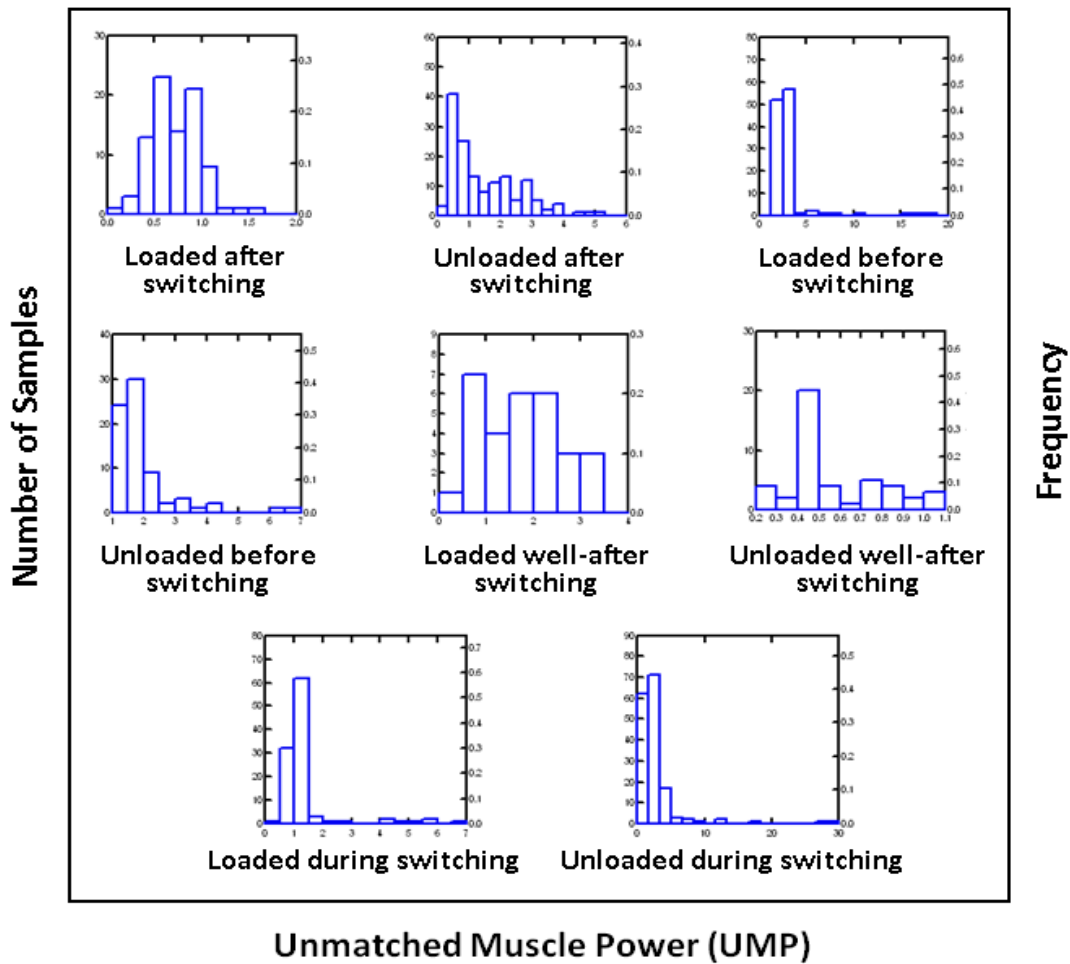

**Figure 8. Histogram for ratio of Amplitude Peaks for Flexor Carpi Radialis (FCR) to MPO measurement for all trials in an experimental block (a.k.a Unmatched Muscle Power). For purposes of analysis, the data were sorted into classes representing intervals of the UMP measurement.**

*Direct effects of loading conditions.* The direct effects of loaded and unloaded conditions can also be compared between the two muscles (TB and FCR). This is the effect on force output and unmatched power during the perturbation itself. The loaded condition resulted in a much lower UMP measurement for the FCR than for the TB. In the case of TB, between 75-80% of the UMP values were between 6 and 14, while none of the values were between 0 and 3. By contrast, 90% of the UMP values generated by the FCR were between 0.5 and 1.5.

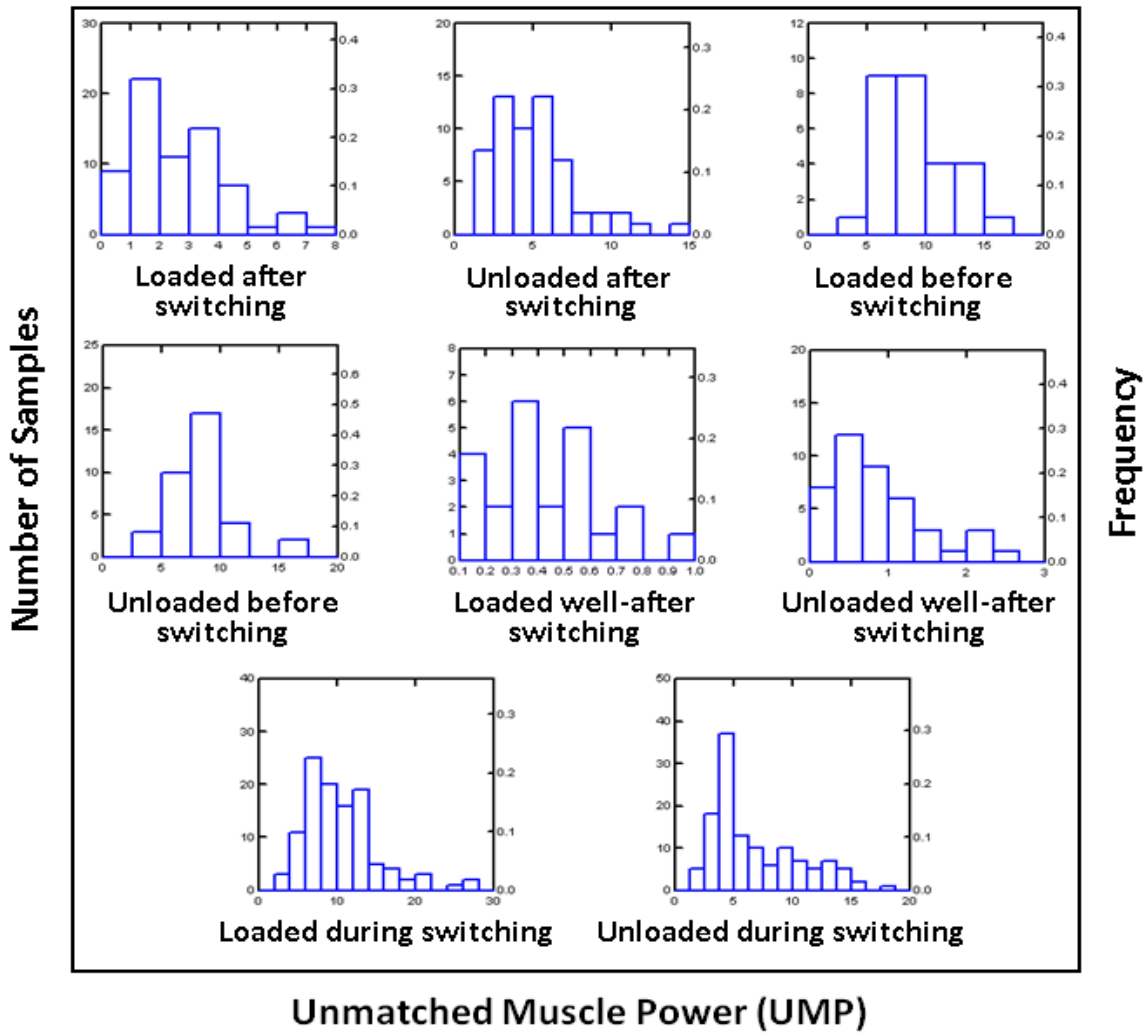

**Figure 9. Histogram for ratio of Amplitude Peaks for Triceps Brachii (TB) to MPO measurement for all trials in an experimental block (a.k.a Unmatched Muscle Power). For purposes of analysis, the data were sorted into classes representing intervals of the UMP measurement.**

These results may mean that TB contributes much more towards overpowered movement during the loaded perturbation, and activity in this muscle may need to be regulated to a greater extent by internal mechanisms after perturbation. A similar but much less pronounced pattern of TB overcompensation is seen in the case of an unloaded perturbation. In addition, UMP measurements using the FCR indicate that an unloaded perturbation results in unpowered movements of the forearm, while UMP measurements using the TB indicate overpowered movements of the humerus.

*Effects well-after initial perturbation.* Finally, it was asked what occurs to the UMP measurement well-after perturbation, or two blocks removed from the original perturbation. In both muscles and for both perturbations, it appears that the UMP measurement increases across the spectrum, with both smaller and larger UMP values being equally represented in the population. This may indicate that changes induced by the perturbation introduced over the course of a single block may not have a lasting effect, at least as it pertains to the matching of required force production to muscle peak amplitude and actual force production.

**Work Loop Trace Analysis**

This analysis was conducted to answer the question of how selected muscles work together to regulate movement. Figures 10, 11, and 13 shows a loop trace of the Triceps Brachii (TB) and Flexor Carpi Radialis muscles for Experiments #1, #2, and #3, respectively. In each graph, the co-contraction of both muscles before, during, and after perturbation is represented. The loop trace is derived from the raw signal, and is a way of demonstrating how the muscle work together functionally across a time-series. Using a graphical representation, we may assess the contributions of two different muscles as they perform work simultaneously. In cases where muscles act synergistically, the EMG trace forms an elliptical trajectory orbiting an attractor point (0,0). Perturbations in the experimental setting will immediately change the loop trace shape, and long-term effects will result in hysteresis (e.g. traces that do not return to its original shape).

*Work loop trace for Experiment #1.* An example from Experiment #1 is shown in Figure 10. For condition 1, when the loaded perturbation is introduced during the learning block 3, the forearm muscle (FCR) temporarily exhibits less amplitude during perturbation. This does not necessarily mean that individuals are compensating by shifting some of the work to the humerus, because there is no corresponding change in the shape of the loop trace. This suggests that internal robustness mechanisms are at work in this case. In condition 3, when the unloaded perturbation is introduced during the learning block 3, the forearm muscle (FCR) exhibits a greater range of amplitude during perturbation. This is in contrast to condition 1, where the humeral muscle (TB) contributes to a slight shift in the loop trace pattern. This may reflect a contribution of morphological control during this set of conditions.

In condition 3, when the loaded perturbation is introduced during the learning block, the extent of the loop trace pattern is limited by the range of the recording equipment. However, the work loop trace shifts inward from these lower and upper bounds in blocks subsequent to the perturbation. This may suggest an interaction between morphological control and internal mechanisms under these conditions. For condition 4, when the unloaded perturbation is introduced during the learning block, the loop trace changes its shape post-perturbation that correspond with changes in amplitude for the humeral muscle (TB). This may suggest a role for internal adaptive mechanisms are at work under these conditions.

*Work loop trace for Experiment #2.* Results for the work loop trace analysis is shown in Figure 11 and Table 5. The lesson of the loop trace from the extended swinging device switching experiment is that control and interleaved conditions result in small but largely insignificant changes in the loop trace. A correlation analysis conducted on the horizontal-vertical, vertical-diagonal, and horizontal-vertical components of loop trace skew revealed that these changes demonstrated similar trends. $H_s$-$V_s$ and $V_s$-$D_s$ values exhibit weak correlations (c-values between .30 and -.30), $V_s$-$D_s$ values exhibit strong negative correlation (c-values below -.65).

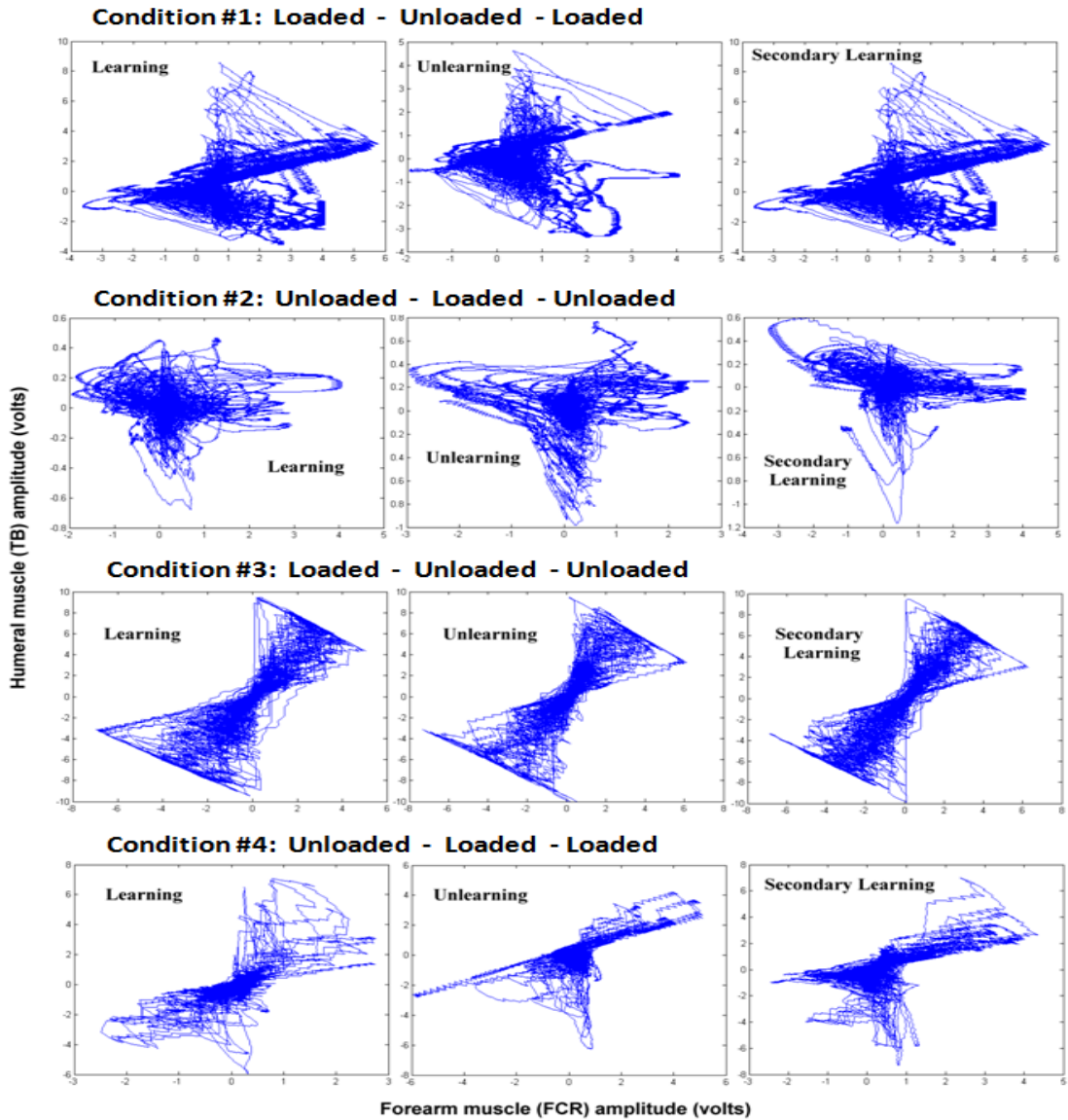

**Figure 10. Loop trace for the prosthetic device switching experiment. The raw signal for the triceps brachii (TB - measured in volts) is plotted against the raw signal for the flexor carpi radialis (FCR – measured in volts). The x-axis represents values attained by the TB muscle, while the y-axis represents values attained by the FCR muscle. Data are arranged in the following manner: each row represents a specific condition, and each column represents a particular set of blocks in the experiment.**

In the interleaved condition, it appears that loading has an effect on $H_s$ and $V_s$ values independent of changes in $D_s$ values. Beyond the finding that uniform (alternating) perturbation and uniform force field conditions yield a similar result, broader relevance of this finding to adaptability and the ability to absorb and learn from the changes in loading remains to be understood.

**Table 5. Percentage of datapoints in each quadrant\* of the work loop trace plot for the extended swinging device switching experiment, by block and condition. Global skew of data points in three directions: horizontally ($H_s$ value), vertically ($V_s$ value), and diagonally ($D_s$ value). Global skew = arbitrary units.**

|  | Block | N, P | P, P | P, N | N, N | $H_s$ value | $V_s$ value | $D_s$ value |
|---|---|---|---|---|---|---|---|---|
| **Interleaved** | L | 0.20 | 0.31 | 0.17 | 0.32 | 0.03 | 0.02 | 0.27 |
|  | U | 0.17 | 0.30 | 0.21 | 0.32 | 0.02 | 0.07 | 0.23 |
|  | SL | 0.16 | 0.34 | 0.26 | 0.24 | 0.21 | 0.00 | 0.16 |
|  | SU | 0.17 | 0.36 | 0.19 | 0.29 | 0.09 | 0.04 | 0.30 |
|  | TL | 0.17 | 0.24 | 0.20 | 0.38 | 0.11 | 0.17 | 0.25 |
| **Early** | L | 0.15 | 0.32 | 0.16 | 0.37 | 0.03 | 0.06 | 0.39 |
|  | U | 0.13 | 0.57 | 0.11 | 0.19 | 0.36 | 0.41 | 0.51 |
|  | SL | 0.16 | 0.39 | 0.17 | 0.28 | 0.11 | 0.10 | 0.35 |
|  | SU | 0.16 | 0.39 | 0.18 | 0.27 | 0.14 | 0.09 | 0.32 |
|  | TL | 0.13 | 0.39 | 0.14 | 0.34 | 0.06 | 0.04 | 0.45 |
| **Late** | L | 0.08 | 0.29 | 0.10 | 0.53 | 0.22 | 0.26 | 0.65 |
|  | U | 0.07 | 0.43 | 0.17 | 0.33 | 0.20 | 0.01 | 0.52 |
|  | SL | 0.05 | 0.41 | 0.07 | 0.47 | 0.03 | 0.09 | 0.76 |
|  | SU | 0.12 | 0.41 | 0.14 | 0.33 | 0.11 | 0.07 | 0.48 |
|  | TL | 0.09 | 0.40 | 0.08 | 0.43 | 0.04 | 0.03 | 0.66 |
| **Control** | L | 0.08 | 0.31 | 0.14 | 0.47 | 0.09 | 0.21 | 0.56 |
|  | U | 0.12 | 0.44 | 0.13 | 0.31 | 0.13 | 0.12 | 0.50 |
|  | SL | 0.15 | 0.36 | 0.25 | 0.25 | 0.21 | 0.02 | 0.21 |
|  | SU | 0.11 | 0.36 | 0.11 | 0.42 | 0.06 | 0.06 | 0.57 |
|  | TL | 0.06 | 0.42 | 0.07 | 0.45 | 0.02 | 0.04 | 0.73 |

\* quadrants of the work loop trace are defined by values along the x- and y-axis in the following way: N, P = negative, positive; P, P = positive, positive; P, N = positive, negative; N, N = negative, negative.

By contrast, conditions that embody early and late loading show changes related to loading, but also changes due to staying in one force field state over three experimental blocks. Using the $D_s$ (diagonal skew) values as a criterion, it appears that the early and late loading conditions exhibit instability within loading perturbations and between loaded and unloaded blocks. However, the control condition shows larger values of $D_s$ in both the first (learning) and final (tertiary learning) block when compared to the third (secondary learning) block. In these cases, a correlation analysis conducted on the horizontal-vertical, vertical-diagonal, and horizontal-vertical components of loop trace skew showed that for the early loading condition, $H_s$, $V_s$, $H_s$, $D_s$, and $V_s$, $D_s$ all have a c-values of above .60 (positive correlation). For the late loading condition, $H_s$, $V_s$ has a c-value of .45 (moderately positive correlation) and $H_s$, $D_s$ has a c-value of -.46 (moderately negative correlation). Changes in the $H_s$ and $V_s$ values between learning and tertiary learning suggests that there is more to be learned with regard to variation in muscle activity during selective perturbation.

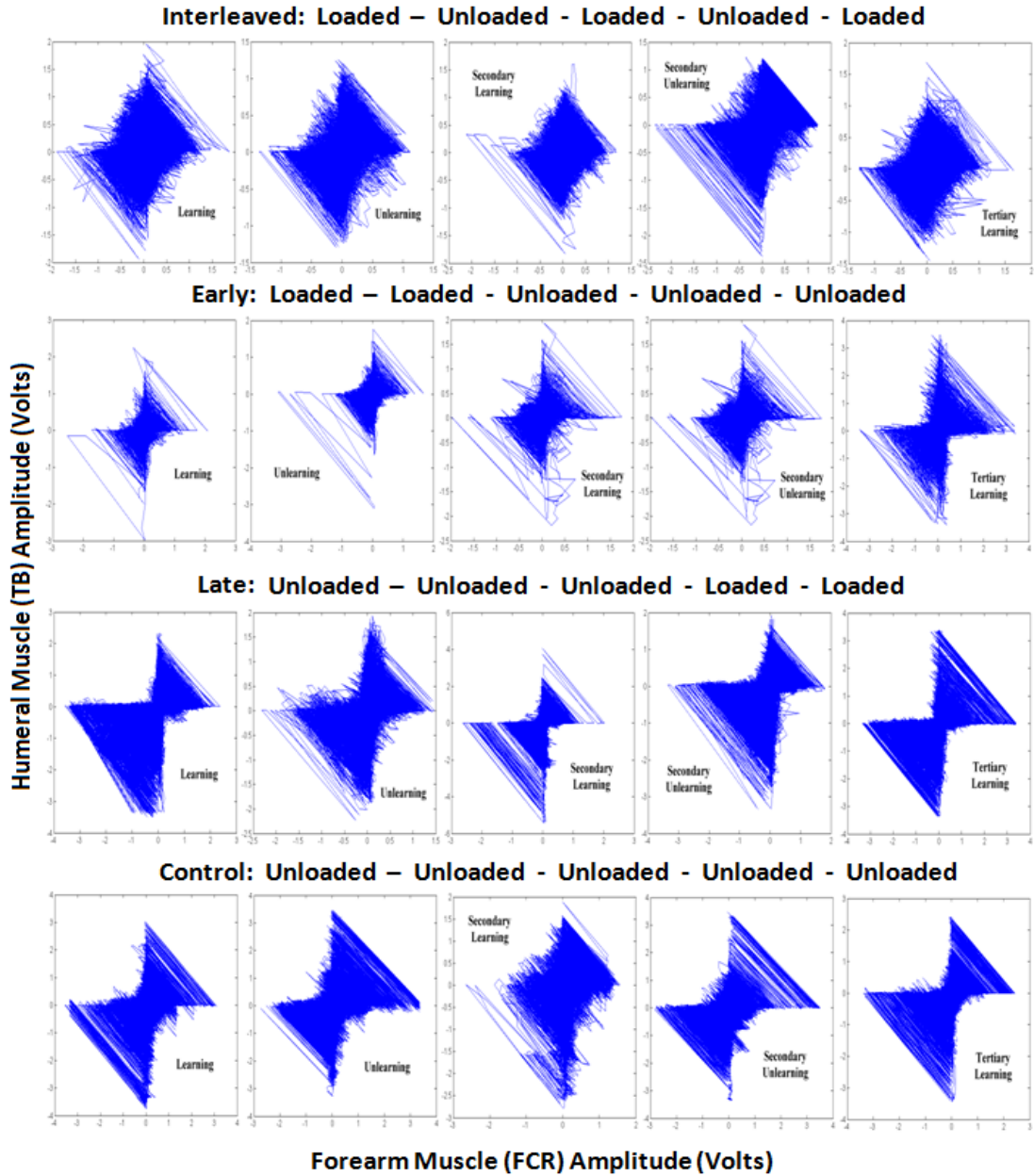

**Figure 11. Work loop trace for the extended prosthetic device switching experiment. The raw signal for the triceps brachii (TB - measured in volts) is plotted against the raw signal for the flexor carpi radialis (FCR – measured in volts). The x-axis represents values attained by the TB muscle, while the y-axis represents values attained by the FCR muscle.**

*4.3.3 Work loop trace for Experiment #3.* An example from the tactile surface switching experiment is shown in Figure 12. For the reverse perturbation condition, there is a change in the loop trace shape both during and after the perturbation. Changes in both the humeral (TB) and forearm muscle (FCR) seem to be independent of each other, although there appears to be a shift in the amplitude of the humeral muscle (TB). This may suggest that slight adjustments are made to muscular control of the humerus during active touch exploration not made during a controlled reaching movement.

For the hard perturbation condition, there is relatively little change in the loop trace shape across the experiment. After perturbation, the amplitude of the forearm muscle (FCR) increases. This is a result similar to the first condition of Experiment 1, and suggests that the same internal regulatory mechanisms are at work. This can be contrasted with the weak perturbation condition, where the loop trace pattern gradually shifts rightward as the amplitude of the humeral muscle (TB) changes. This change occurs during and after perturbation, which suggests either an internal adaptive mechanism or an interaction between internal adaptive mechanisms and morphological control.

## DISCUSSION

One interesting result of these experiments was that a perturbation, whether it was of a different type or at a different position in the condition, has a different effect on physiological output and muscle activity. This can be explained theoretically by further considering the mechanisms of internal, morphological, and environmental regulation and how they relate to the capacity for adapting to changes in the environment. In addition, the results show that there may be two separate regulatory mechanisms: one governing the retention of temporal information about the perturbation, and another that exploits information inherent in the haptic/proprioceptive environment.

**Learning and Physiological Regulation**

One way to connect theoretical concerns to applications is by considering the relationship between learning and physiological regulation [28]. In this case, "regulation" involves how haptic/proprioceptive sensory information is processed, selected upon, and represented in both the peripheral and central nervous systems [29]. When presented with an environmental challenge, a participant will respond in one of two ways. The first way is to exhibit performance decrements characterized by work loop trace skew and large amounts of unmatched muscle power. This represents a failure to adapt. The second way is to be robust or become increasingly so to perturbations over time. This is a much more subtle response, and most likely involves some components of procedural and/or motor learning [23, 11].

*Components of the adaptive response.* The adaptive response itself has two components: innate and acquired. The innate response is related to physiological state, and is related to abilities involving motor coordination and muscle strength. Related work on this same dataset [20, 30] and the literature on prosthetics engineering [31] suggest that limb size and shape may contribute to differences in performance. Based on work using virtual reality to restore balance and enable movement rehabilitation [32, 33], it is also predicted that potential applications to clinical populations will need to take context- and patient-specificity into account.

On the other hand, while the learning response may be stratified by physical ability, the juxtaposition of differential forces may allow for invariant features in both sets to be more readily extracted and applied to future contexts [34]. The extraction of these common features by the nervous system should be highly similar across individuals. They are driven by the structure of haptic/proprioceptive sensory information, thus comprising an important component of any potential application.

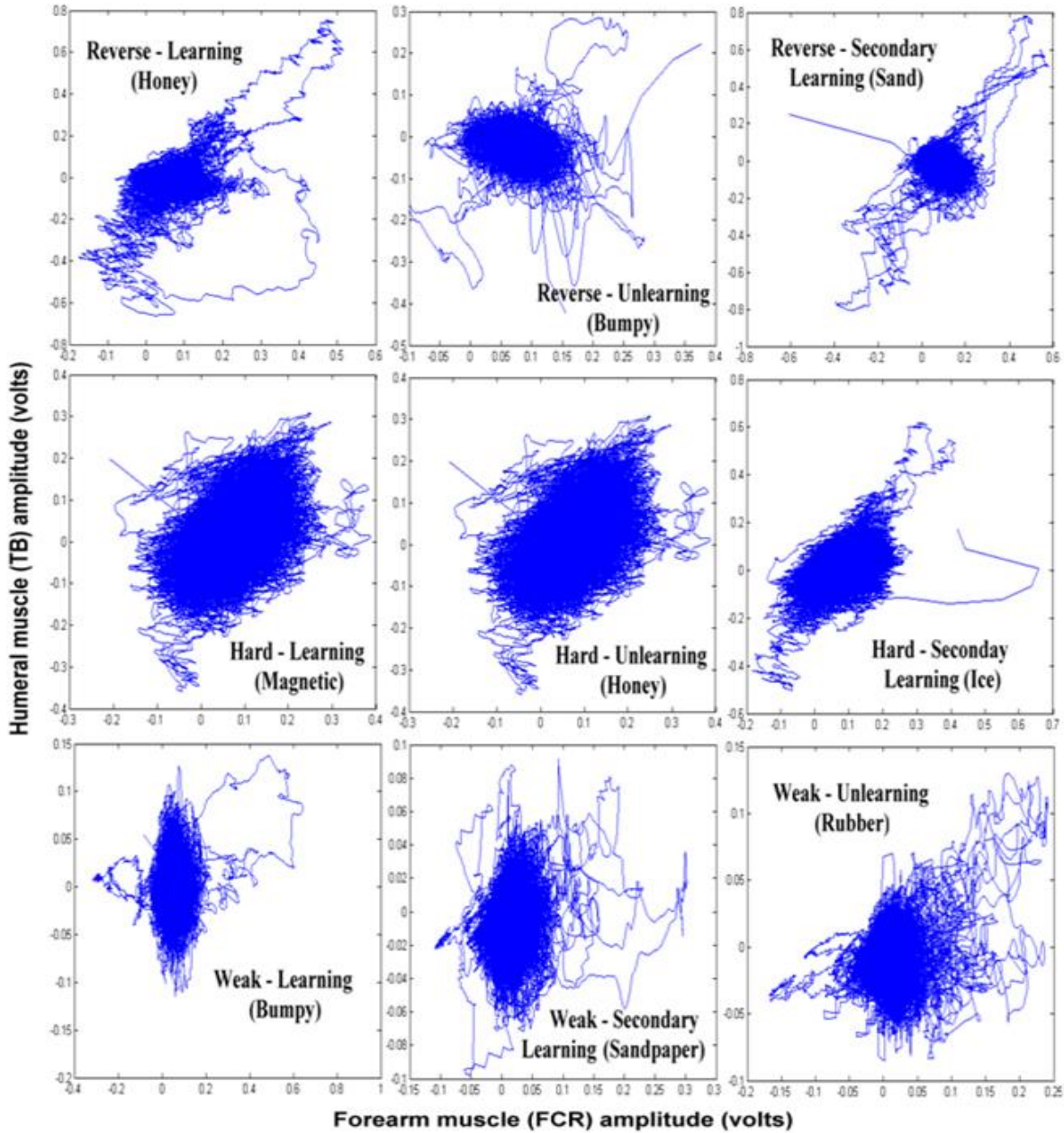

**Figure 12. Loop trace for the tactile surface resistance switching experiment. The raw signal for the triceps brachii (TB - measured in volts) is plotted against the raw signal for the flexor carpi radialis (FCR – measured in volts). The x-axis represents values attained by the TB muscle, while the y-axis represents values attained by the FCR muscle.**

**Learning and Physiological Adaptation as Regulatory Mechanisms**

The right sequence of perturbations may play a particularly important role in reinforcing environmental features that contain information. In the physiological context, the concepts of homeostasis and allostatic drive [35] may ultimately be critical to understanding how technologies, from touch-driven interfaces to motion-driven virtual environments, interact with a person's physiology and elicit a particular response. It is worth noting that the consensus in brain-computer interface (BCI) design [36, 37] does not directly address this issue. Whether perceptual cues provided by a selectively distorted environment are context-specific or related to a specific set of muscles and brain regions remains to be seen.

*Applications to Manipulable Technologies.* Manipulable technologies, or information technologies that you can touch, move, and otherwise physically control, can play a role in both medical and non-medical applications. In the realm of medicine, customizable manipulable interfaces may aid people who suffer from movement disorders and aging-related muscle weakness. Manipulable interfaces can function as both training devices for rehabilitation and as aids for making otherwise hard-to-manipulate devices more usable. For non-medical applications, the work presented in this paper might be adapted to training people of various anthropometric dimensions to use touch screens of different sizes and degrees of capacitance. The goal for future medical and non-medical applications alike should revolve around optimizing which extremes and patterns that define certain sets of forces (e.g. pendular-like reaching movements or ballistic exploratory movements) are most helpful in the recovery of function and maintenance of performance, respectively.